\documentclass[cameraready]{Interspeech}
\title{Controllable Accent Normalization via Discrete Diffusion}

\author[affiliation={1,5}, orcid=0009-0000-2812-9197]{Qibing}{Bai}
\author[affiliation={4}, orcid=0009-0007-8278-031X]{Yuhan}{Du}
\author[affiliation={6}, orcid=0000-0002-5324-8961]{Tom}{Ko}
\author[affiliation={4,6}, orcid=0000-0003-1523-9631, correspondingauthor]{Shuai}{Wang}
\author[affiliation={5}, orcid=0009-0001-9062-513X]{\\Yannan}{Wang}
\author[affiliation={2,3,6}, orcid=0000-0001-9158-9401, correspondingauthor]{Haizhou}{Li}

\address{
    $^1$ SDS, $^2$ SAI, and $^3$ SRIBD, The Chinese University of Hong Kong, Shenzhen, China \\
    $^4$ School of Intelligence Science and Technology, Nanjing University, Suzhou, China \\
    $^5$ Tencent Ethereal Audio Lab, Tencent, Shenzhen, China \\
    $^6$ Shenzhen Loop Area Institute, Shenzhen, China
}

\email{qibingbai@link.cuhk.edu.cn, shuaiwang@nju.edu.cn, haizhouli@cuhk.edu.cn}


\usepackage{comment}
\usepackage{multirow}
\usepackage{algorithm}
\usepackage{algpseudocode}
\usepackage{cite}

\begin{document}

\maketitle

\begin{abstract}
  Existing accent normalization methods do not typically offer control over accent strength, yet many applications---such as language learning and dubbing---require tunable accent retention. We propose DLM-AN, a controllable accent normalization system built on masked discrete diffusion over self-supervised speech tokens. A Common Token Predictor identifies source tokens that likely encode native pronunciation; these tokens are selectively reused to initialize the reverse diffusion process. This provides a simple yet effective mechanism for controlling accent strength: reusing more tokens preserves more of the original accent. DLM-AN further incorporates a flow-matching Duration Ratio Predictor that automatically adjusts the total duration to better match the native rhythm. Experiments on multi-accent English data show that DLM-AN achieves the $\text{lowest}$ word error rate among all compared systems while delivering competitive accent reduction and smooth, interpretable $\text{accent strength control}$\footnote{Samples: https://P1ping.github.io/dlman-demo/}. The implementation is available at \url{https://github.com/P1ping/DLM-AN}.
\end{abstract}

\section{Introduction}
Accent conversion (AC) seeks to alter speech from one accent to another while preserving the speaker's characteristics. A special case, accent normalization (AN)\footnote{Also referred to as foreign accent conversion (FAC).}, converts non-native (L2) accented speech into a native (L1) accented form. AN technology enables a wide range of applications, including pronunciation training for language learners~\cite{felps2009foreign}, authentic dubbing in multimedia~\cite{turk2002subband}, and personalized text-to-speech systems~\cite{sun2016personalized}.

Early deep learning approaches for AN are \textit{reference-based}~\cite{zhao2018icassp, zhao2019foreign, li2020improving, ding2022accentron}, relying on native accent speech samples to generate accent-neutral representations via PPG features~\cite{zhao2018icassp, zhao2019foreign, ding2022accentron} or native TTS~\cite{li2020improving}. \textit{Reference-free methods}~\cite{zhao2021converting, nguyen2022accent, quamer2022zero} eliminate this requirement by directly mapping between accented and native representations, though they still rely on parallel data. Subsequent work removes the parallel data constraint through ASR--TTS pipelines~\cite{liu2020end}, accent feature disentanglement~\cite{jin2023voice}, or TTS-guided representations~\cite{zhou2023tts, chen2024transfer}, further augmented with flow matching~\cite{bai2024diffusion} and normalizing flow~\cite{nguyen2024improving}. However, these approaches depend on TTS-synthesized targets, whose quality can be limited by voice cloning and duration modeling errors.

Recent token-based methods~\cite{nguyen24syndata4genai,jia2024convert,bai2025accent,zhang2025vevo} offer a promising alternative. TokAN~\cite{bai2025accent} quantizes speech into self-supervised discrete tokens, performs autoregressive token-to-token conversion, and recovers waveforms via a flow-matching synthesizer. CosyAccent~\cite{bai2026cosyaccent} adopts a non-autoregressive direct flow-matching approach and introduces a ``source-synthesis'' data strategy. Both systems support total-duration control. However, none of the above methods provides control over accent strength---a desirable capability for gradual accent reduction or accent preservation as part of speaker identity. The only attempt at accent strength control~\cite{halychanskyi2025fac} manipulates the starting timestep of a continuous diffusion process, but it operates under a frame-to-frame framework with fixed duration, lacking fine-grained rhythm adjustability and duration control.

In this paper, we propose \textbf{DLM-AN} (\textbf{D}iffusion \textbf{L}anguage \textbf{M}odel for \textbf{A}ccent \textbf{N}ormalization), a controllable accent normalization system based on masked discrete diffusion over self-supervised speech tokens. DLM-AN extends the LLaDA diffusion language model~\cite{nie2025llada} to speech: a bidirectional Transformer iteratively predicts masked tokens conditioned on content representations from a CTC-guided~\cite{graves2006connectionist} token encoder. The key insight enabling controllability is that, under a phonetically rich tokenizer, native and accented renditions of the same utterance share many tokens in similarly pronounced regions but differ in accent-affected regions. We introduce a Common Token Predictor (CTP) that identifies tokens likely shared with the native target. By reusing high-confidence source tokens to initialize the masked sequence, users can smoothly control accent strength---from full normalization (all tokens generated from scratch) to near-resynthesis (all source tokens preserved). A flow-matching Duration Ratio Predictor further provides explicit control over the total output duration.

Our contributions are as follows:
\begin{itemize}
\item We propose the first accent normalization system based on discrete diffusion, enabling iterative token generation conditioned on phonemically guided content representations.
\item We introduce a common token predictor that provides smooth, interpretable accent strength control through threshold-based source token reuse.
\item We demonstrate that DLM-AN achieves the best content preservation (lowest WER) among all compared systems, while offering competitive naturalness and accent reduction, along with robust duration scaling.
\end{itemize}

\section{Related Work}
\subsection{Controllability in Accent Conversion}
Most accent conversion/normalization systems perform a one-shot ``full'' accent shift without a user-controllable knob~\cite{zhao2019foreign,zhang2025vevo}. Recently, controllability has attracted increasing attention, motivated by applications such as gradual accent reduction in language learning and adjustable accent retention in dubbing.

One line of work studies duration control (speech-rate control) during accent normalization. TokAN~\cite{bai2025accent} performs token-to-token conversion with self-supervised discrete units and supports duration preservation. CosyAccent~\cite{bai2026cosyaccent} further enables explicit duration scaling via a duration-ratio predictor and proposes a source-synthesis data strategy to reduce reliance on TTS-synthesized supervision artifacts.

In contrast, controllable 
\emph{accent strength} (i.e., smoothly trading off normalization vs.\ retaining the original L2 accent) remains less explored. FAC-FACodec~\cite{halychanskyi2025fac} introduces an intensity control mechanism in a diffusion-based, factorized-codec framework by using the diffusion starting timestep as a user knob. Related directions include fine-grained controllable accent transfer models~\cite{wang2023nonparallel}, controllable accented TTS that renders accent intensity at coarse/fine levels~\cite{liu2024controllable}, and scalable accented TTS with automated accent label discovery~\cite{xinyuan2025scalable}.

Our work leverages masked discrete diffusion over speech tokens to support both total-duration control and an interpretable accent-strength knob via source-token reuse; the iterative masking/unmasking procedure also naturally supports localized correction (speech infilling).

\subsection{Self-Supervised Speech Tokens}
Discrete tokens derived from self-supervised learning (SSL) representations~\cite{hsu2021hubert,chen2022wavlm} correlate strongly with phonetic content~\cite{choi24self}, which makes them attractive units for speech generation and conversion. Such tokens have been adopted in token-based voice conversion~\cite{huang2021any,kreuk2022textless,oh2025durflex} and high-fidelity speech generation/TTS frameworks~\cite{kharitonov2023speak}. More broadly, discreteness allows importing ``text-like'' modeling techniques into speech, including spoken language modeling~\cite{lakhotia2021generative}, direct speech-to-speech translation~\cite{lee2022direct}, and speech-centric LLM systems~\cite{fang2024llamaomni}.

Recent work also explores SSL tokens for ASR, including multilingual ASR~\cite{cui2025exploring1}, contextual ASR~\cite{cui2025exploring2}, and accent-robust discrete-token ASR modeling~\cite{onda2026advanced}. For accent conversion/normalization, discrete tokens have been studied under different supervision regimes, including zero-shot/minimally-supervised conversion~\cite{jia2024convert}, pseudo-parallel mapping~\cite{bai2025accent,nguyen24syndata4genai}, and prompt-based imitation~\cite{zhang2025vevo}.
We leverage WavLM~\cite{chen2022wavlm} to extract discrete tokens for conversion and synthesis, capitalizing on its strong phonetic encoding and noise robustness.

\subsection{Discrete Diffusion}
\label{sec:discrete_diffusion}
Discrete diffusion adapts diffusion generative modeling to categorical data such as tokens, and has recently been investigated for language modeling. Two commonly used formulations are: (i) uniform/multinomial diffusion, where tokens transition probabilistically across the vocabulary~\cite{hoogeboom2021argmax,austin2021structured}, and (ii) absorbing/masked diffusion, where tokens are progressively mapped to a special absorbing state (e.g., \texttt{[MASK]}) and then iteratively recovered by a learned reverse process~\cite{austin2021structured,sahoo2024simple,nie2025llada}.

We focus on the absorbing formulation, which underlies recent diffusion language models such as LLaDA~\cite{nie2025llada}. Let \(\mathbf{x}_0 \in \{1,\dots,V\}^n\) be a clean token sequence with vocabulary size \(V\) and length \(n\), and let \(m\) denote the absorbing mask token. We parameterize the diffusion in continuous time \(t \in [0,1]\) with a differentiable non-increasing survival schedule \(\alpha(t)\), where \(\alpha(0)\approx 1\) and \(\alpha(1)\approx 0\). The forward marginal independently corrupts each token as
\begin{equation}
q(\mathbf{x}_t \mid \mathbf{x}_0) =
\prod_{i=1}^n
\mathrm{Cat}\left(
x_t^i;\,
\alpha(t)\mathbf{e}_{x_0^i}
+
(1-\alpha(t))\mathbf{e}_{m}
\right),
\end{equation}
i.e., each token is kept clean with probability \(\alpha(t)\) or absorbed into \(m\) with probability \(1-\alpha(t)\). Once masked, a token remains masked in the forward process, and \(\mathbf{x}_1\) is almost fully masked.

The reverse process is commonly parameterized through an \(x_0\)-prediction network, typically a bidirectional Transformer that predicts clean-token distributions in parallel. For \(0\leq s<t\leq 1\), the SUBS parameterization~\cite{sahoo2024simple} substitutes the predicted clean sequence distribution into the analytic absorbing-diffusion posterior. Let \(\boldsymbol{\pi}_\theta^i(\mathbf{x}_t,t)=p_\theta(x_0^i=\cdot\mid \mathbf{x}_t)\) denote the predicted clean-token distribution at position \(i\). Define \(\kappa_{s,t}=(\alpha(s)-\alpha(t))/(1-\alpha(t))\) and \(\bar{\kappa}_{s,t}=(1-\alpha(s))/(1-\alpha(t))\). With $\boldsymbol{\mu}_{s,t}^i = \bar{\kappa}_{s,t}\mathbf{e}_{m} + \kappa_{s,t}\boldsymbol{\pi}_\theta^i(\mathbf{x}_t,t)$, the reverse process is
\begin{equation}
\begin{aligned}
p_\theta(x_s^i \mid \mathbf{x}_t)
&=
\left\{
\begin{array}{@{}ll@{}}
\mathrm{Cat}(x_s^i;\mathbf{e}_{x_t^i}),
&\! x_t^i \neq m, \\[1mm]
\mathrm{Cat}(x_s^i;\boldsymbol{\mu}_{s,t}^i),
&\! x_t^i = m,
\end{array}
\right.
\end{aligned}
\label{eq:absorbing_reverse}
\end{equation}
SUBS enforces two absorbing-process properties: zero probability of predicting the mask token, and carry-over unmasking, where visible tokens are copied unchanged during reverse diffusion. These substitutions simplify the discrete-time D3PM~\cite{austin2021structured} negative ELBO; taking the continuous-time limit further yields the simplified MDLM objective~\cite{sahoo2024simple}, in which only corrupted positions contribute to the effective denoising loss. The continuous-time likelihood-bound objective becomes a reweighted masked-token cross-entropy:
\begin{equation}
\mathcal{L}(\theta)
=
\mathbb{E}_{t,\mathbf{x}_0,\mathbf{x}_t}
\left[
\frac{-\alpha'(t)}{1-\alpha(t)}
\sum_{i:\, x_t^i=m}
-\log p_\theta(x_0^i \mid \mathbf{x}_t)
\right],
\label{eq:dlm_general}
\end{equation}
where $t \sim \mathcal{U}(0, 1]$, $\mathbf{x}_t \sim q(\cdot | \mathbf{x}_0)$, and the expectation is approximated by Monte Carlo sampling. Although the MDLM objective is often written as an all-token loss, visible tokens do not contribute to the loss in the carry-over setting, yielding the masked-position form above. The weight \(-\alpha'(t)/(1-\alpha(t))\) is the instantaneous reverse unmasking rate; for the linear survival schedule \(\alpha(t)=1-t\), it reduces to \(1/t\), matching the LLaDA-style objective used in our methodology. In practice, \(t\) is bounded away from \(0\) for numerical stability.

Compared with heuristic masked iterative generators~\cite{chang2022maskgit,wang2025maskgct,wang2025metis}, masked diffusion specifies an explicit forward corruption process and optimizes a likelihood-bound objective while retaining parallel iterative denoising. Surveys highlight its growing role in LLMs~\cite{yu2025discrete}. Recent work further explores discrete-score formulations~\cite{lou2024discrete}, corrective/remasking models~\cite{wang2025remasking,zhang2026corrective,song2025seed,bie2026llada2} and efficient samplers~\cite{benhamu2025accelerated}.

\section{Methodology}

\begin{figure*}[t]
\centering
\includegraphics[width=0.95\textwidth]{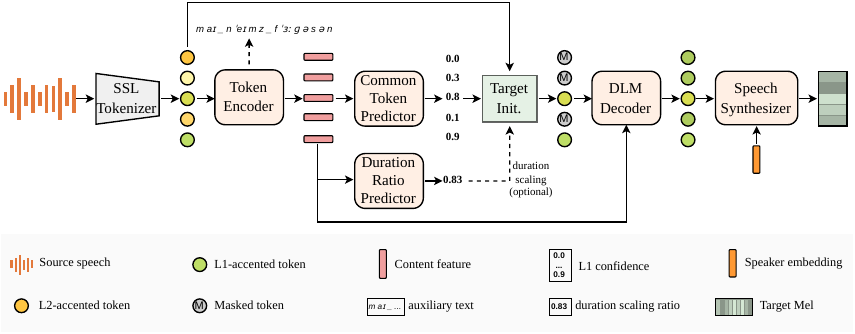}
\vspace{-0.3cm}
\caption{Overview of the DLM-AN pipeline. The SSL tokenizer extracts discrete tokens from L2-accented speech. A Transformer token encoder with CTC-based phonemic guidance produces content representations, which are fed into the Common Token Predictor (CTP), Duration Ratio Predictor (DP), and the DLM decoder. The DLM decoder iteratively generates the target token sequence, optionally initialized with high-CTP-confidence source tokens. A flow-matching synthesizer and vocoder produce the final waveform.}
\vspace{-0.5cm}
\label{fig:overview}
\end{figure*}

We propose to use a diffusion language model (DLM) for controllable accent normalization. The method is shortened ``DLM-AN". Figure~\ref{fig:overview} shows the pipeline of DLM-AN. The SSL tokenizer extracts SSL representations from the L2-accented waveform and quantize the features into discrete tokens. A Transformer token encoder takes further processs these tokens, producing continuous content representations. To make the content representations phonemic enough, a CTC-based phonemic guidance is imposed upon them, shown as the auxiliary text in Figure~\ref{fig:overview}. The content representations are further taken as the input for three modules: Common Token Predictor (CTP), Duration Ratio Predictor (DP), and the DLM decoder.

Note that the content features are concatenated with the source tokens (embeddings) when feeding to CTP and DP, in order to provide detailed pronunciation patterns (i.e., phonetic information). This concatenation is not displayed in Figure~\ref{fig:overview} for brevity.

CTP predicts whether each token is common in both the source and target sequences. The higher the score for a token, the more probable this token is associated with a native pronunciation, as will be demonstrated in the subsequent sections. Tokens with high CTP scores could be optionally ``reused" for the decoding/generation process later. DP predicts the ratio of the total duration: $\text{dur}_\text{tgt} / \text{dur}_\text{src}$. This ratio can be optionally used to determine the total duration/length of the target sequence.

Given the target duration ratio, either predicted, arbitrarily specified, or kept 1.0 to maintain the source duration, the length of the initial target sequence is determined. By default, the initial target sequence is purely filled with \texttt{[MASK]} for generation from scratch. Optionally, based on the CTP scores and a given threshold or proportion, certain soruce tokens can be reused to initialize the target sequence. After the initial target sequence is determined, the DLM decoder iteratively generates the entire target sequence, conditioned on the content features from the token encoder.

The more tokens reused in the target sequence, the more source accent is expected to be preserved. The speech synthesizer further generates the corresponding Mel-spectrogram given the target tokens, conditioned on the speaker embedding extracted from the input source speech. The Mel-spectrogram can be converted to a waveform using the HiFT vocoder~\cite{li2023hiftnet}.

\subsection{Discrete Diffusion for Speech Tokens}

We extend the LLaDA masked diffusion language model~\cite{nie2025llada} to discrete speech tokens for controllable accent normalization. Our forward corruption process uses pure absorbing masking, parameterized by a timestep \(t \sim \mathcal{U}[0,1]\) per sequence.

Let \(\mathbf{y}_0 \in \{1,\dots,V\}^L\) be a clean speech token sequence, where \(V\) is the speech vocabulary size and \(L\) is the sequence length. Following the notation in Section~\ref{sec:discrete_diffusion}, we deploy $\alpha(t) = 1 - t$ and denote the masking probability as \(\lambda(t)=1-\alpha(t)\). We use the numerically stable linear form \(\lambda(t)=(1-\epsilon)t+\epsilon\), with \(\epsilon>0\). Each position \(i\) is masked independently:
\begin{align}
q_{\lambda}(z_i \mid y_0^i)
&= \lambda(t)\,\delta_{z_i=\texttt{[MASK]}} + (1-\lambda(t))\,\delta_{z_i=y_0^i}
\end{align}
which induces a masked index set \(M\) and a visible set \(\bar{M} = [L]\setminus M\).
The corrupted sequence \(\mathbf{z}\) has corruption set \(\mathcal{C} = M\). The model \(p_\theta(\mathbf{y}_0 \mid \mathbf{z}, \mathbf{c})\) is a bidirectional Transformer that predicts the original tokens from \(\mathbf{z}\), conditioned on the content representations \(\mathbf{c}\) from the token encoder. The training objective follows the LLaDA-style masked-position loss:
\begin{align}
\mathcal{L}_{\text{DLM}}(\theta)
&=
-\mathbb{E}_{t,\mathbf{y}_0,\mathbf{z}}
\left[
\frac{1}{\lambda(t)}
\sum_{i \in M}
\log p_\theta(y_0^i \mid \mathbf{z}, \mathbf{c})
\right],
\label{eq:dlm_loss}
\end{align}
where \(\mathbf{z}\sim q_{\lambda(t)}(\cdot\mid \mathbf{y}_0)\). The expectation is approximated by Monte Carlo sampling with global per-token normalization for stability. This objective is the masking-rate form of the likelihood-bound objective in Eq.~\eqref{eq:dlm_general}; the small \(\epsilon\) only bounds the weighting near \(t=0\).

The DLM decoder has a decoder-only Transformer structure without causal masking, allowing parallel prediction over masked positions. The content representations \(\mathbf{c}\) are integrated as conditional inputs via self-attention to guide accent-normalized generation while preserving content.

During inference, we employ a greedy sampling algorithm with optional initialization: start from a fully or partially masked sequence (reusing high-CTP-confidence tokens), and iteratively predict and unmask positions based on confidence scores, conditioned on \(\mathbf{c}\). For enhanced quality, we use classifier-free guidance (CFG)~\cite{ho2021classifierfree,nie2025llada}, combining the conditional and unconditional predictions at the logit level. Let \(\boldsymbol{\ell}_\text{cond}\) and \(\boldsymbol{\ell}_\text{uncond}\) be the logits corresponding to \(p_\theta(\mathbf{y}_0\mid\mathbf{z},\mathbf{c})\) and \(p_\theta(\mathbf{y}_0\mid\mathbf{z},\varnothing)\), respectively. The guided distribution is then obtained as $p_{\mathrm{cfg}}(\cdot) = \operatorname{softmax}\left((1+w_{\text{DLM}})\boldsymbol{\ell}_\text{cond} - w_{\text{DLM}}\boldsymbol{\ell}_\text{uncond}\right)$, where \(w_{\text{DLM}}\) controls the guidance strength.

\begin{figure}[htbp]
\centering
\vspace{-0.2cm}
\includegraphics[width=0.35\textwidth]{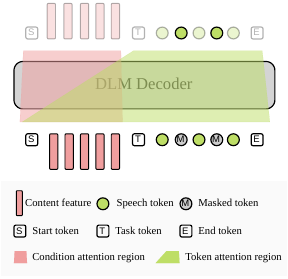}
\vspace{-0.3cm}
\caption{Structure of the DLM decoder. The input consists of the content features and the noised target tokens, separated by special tokens \texttt{[START]}, \texttt{[TASK]}, and \texttt{[END]}. The content features are mutually attentive but do not attend to the token sequence (pink region), while the token sequence attends to the entire input (green region).}
\label{fig:dlm}
\vspace{-0.3cm}
\end{figure}

The structure of DLM decoder is shown in Figure~\ref{fig:dlm}. It features a self-attention-only structure (i.e., using self-attention to fuse the conditional information). The input is basically two sequences: the content features and the noised target tokens. Three special tokens, \texttt{[START]}, \texttt{[TASK]}, and \texttt{[END]}, are added to wrap and separate the two sequences, guiding the DLM behavior.

To make the condition reusable across time steps (i.e., token sequences with different numbers of masked tokens), the content features are mutually attentive but with no attention to the token sequence. In contrast, the token sequence can attend to the entire input sequence. The attention regions of the two parts are depicted in Figure~\ref{fig:dlm} with the colors pink and green, respectively.

\subsection{Common Token Prediction}
With a sufficiently phonetic tokenizer, utterances spoken with different accents tend to share many tokens in similarly pronounced regions, while differing mainly in accent-affected regions. Motivated by this property, we introduce a Common Token Predictor (CTP) that assigns each source token a confidence score indicating how likely it is to be shared with the (native) target. Tokens with high CTP confidence can be reused to initialize the target sequence, providing a simple and interpretable control of accent strength: reusing more tokens preserves more of the source accent.
\begin{figure}[htbp]
\centering
\vspace{-0.3cm}
\includegraphics[width=0.45\textwidth]{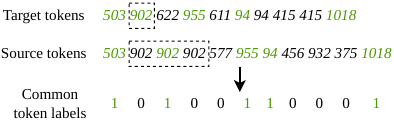}
\caption{Extraction of common token labels via the longest common subsequence (LCS) between source and target token sequences. For consecutive identical tokens with differing durations, center-mode alignment is applied (dashed rectangle).}
\vspace{-0.2cm}
\label{fig:lcs_computation}
\end{figure}
We formulate CTP as a sequence-tagging problem. Given paired source and target token sequences, we derive binary labels by computing the longest common subsequence (LCS) between them. LCS has been used previously to evaluate accent conversion models~\cite{jia2024convert}; here we use it to identify which source tokens are shared with the target. The label extraction procedure is illustrated in Figure~\ref{fig:lcs_computation}. We obtain the LCS via dynamic programming and backtracking. For consecutive identical tokens that have different durations in the source and target, we apply a center alignment: only the centered tokens are marked positive (dashed rectangle in Figure~\ref{fig:lcs_computation}).

The training objective for CTP is binary cross-entropy loss. Let \(S\) be the source sequence length, and let \(\mathbf{o} \in \{0,1\}^S\) be the binary labels derived from the LCS searching. Given the source content features, the CTP module (parameterized by \(\phi\)) outputs predicted probabilities \(\hat{\mathbf{o}} \in [0,1]^S\). The loss is:
\begin{align}
\mathcal{L}_{\text{CTP}}(\phi)
&= -\frac{1}{S} \sum_{i=1}^S \left[ o_i \log \hat{o}_i + (1 - o_i) \log (1 - \hat{o}_i) \right].
\end{align}

\subsection{Duration Ratio Prediction}
Because L2-accented speech often exhibits different rhythm and speaking rate, directly inheriting the source total duration can lead to sub-optimal naturalness. DLM-AN therefore includes a duration ratio predictor (DP) that estimates the global duration ratio $r=\text{dur}_{\text{tgt}}/\text{dur}_{\text{src}}$. DP uses a diffusion Transformer (DiT)~\cite{peebles2023scalable} backbone with an attentive pooling layer, and is trained with conditional flow matching.

The training objective for DP is conditional flow matching loss~\cite{lipman2023flow}. Let \(r > 0\) be the target duration ratio (ground truth \(\text{dur}_{\text{tgt}} / \text{dur}_{\text{src}}\)), and let \(\mathbf{c}\) be the input content representations from the token encoder. The flow matching model \(v_{\psi}(u_t, t, \mathbf{c})\) (parameterized by \(\psi\)) predicts the velocity field:
\begin{align}
u_t &= (1-t)\, u_0 + t\, r,
\end{align}
where $t \sim \mathcal{U}[0,1]$ and $u_0 \sim \mathcal{N}(0,1)$ is a standard normal prior. The loss is:
\begin{align}
\mathcal{L}_{\text{DP}}(\psi)
&= \mathbb{E}_{t,\, u_0} [ \Vert v_{\psi}(u_t, t, \mathbf{c}) - (r - u_0) \Vert^2 ]
\end{align}
This objective trains the model to generate duration ratios conditioned on \(\mathbf{c}\), enabling global rhythm adjustment. In practice, we also condition DP on the source token embeddings to capture fine-grained pronunciation details; we omit this from the formulation for brevity.

\begin{figure*}[htbp]
\centering
\includegraphics[width=0.95\textwidth]{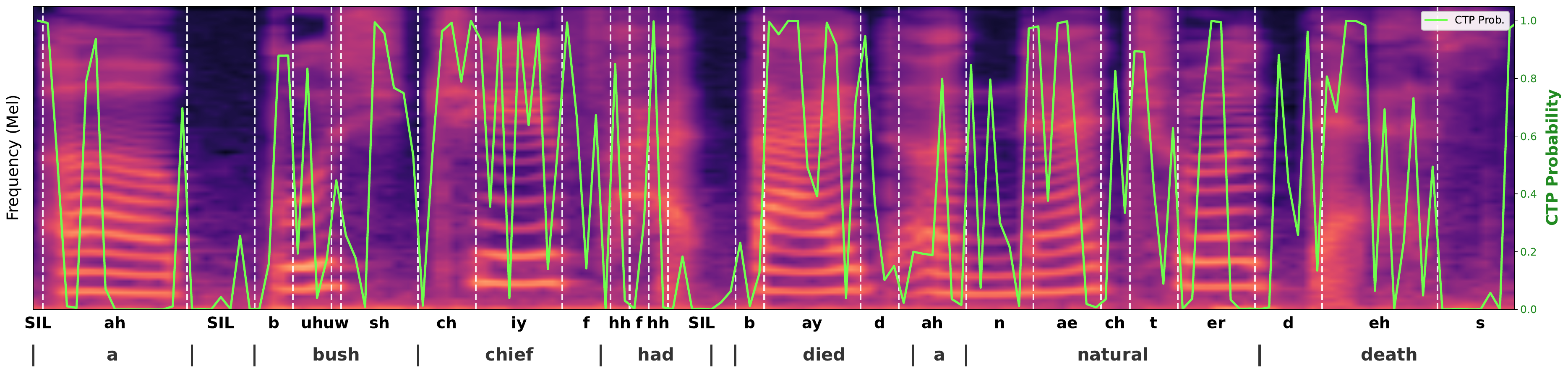}
\vspace{-0.2cm}
\caption{Visualization of common token prediction for a Chinese-accented sample. CTP confidence values are overlaid on the Mel-spectrogram. PPG-predicted phonemes are shown below and their boundaries (white dashed lines) are overlaid on the spectrogram. Aligned words are shown at the bottom. Regions with prominent L2 accent (e.g., prolonged ``a'', unclear ``had'', /S/-like ending of ``death'') receive low CTP confidence.}
\label{fig:lcs_vis}
\vspace{-0.4cm}
\end{figure*}

\subsection{Phoneme Guidance and Joint Training}

The token encoder is a Transformer with relative positional embeddings, producing content representations \(\mathbf{c}\). The CTP and DP modules also use Transformer backbones with relative positional embeddings. The DLM decoder is a self-attention-only Transformer with two-block masking, using rotary positional encoding (RoPE)~\cite{su2024roformer} for the entire sequence.

To encourage \(\mathbf{c}\) to be phonemically informative, we impose a CTC-based phonemic guidance on the token encoder outputs. Concretely, we attach a linear projection head on top of the token encoder to predict phoneme logits, and compute a CTC loss against phoneme label sequences derived from the corresponding transcripts. Given phoneme sequence $\mathbf{p}$, the loss is
\begin{align}
\mathcal{L}_{\text{CTC}} &= \mathrm{CTC}(\text{Linear}(\mathbf{c}),\, \mathbf{p})
\end{align}

We train the token-prediction modules jointly, including the token encoder, CTP, DP, and DLM decoder, following a two-stage schedule of pretraining and fine-tuning as described in the experimental setup. Let \(\mathcal{L}_{\text{DLM}}\) denote the masked discrete diffusion loss for token generation (Eq.\,(\ref{eq:dlm_loss})). The overall joint-training objective is:
\begin{align}
\mathcal{L}_{\text{token}} = \mathcal{L}_{\text{DLM}} + \beta_1\mathcal{L}_{\text{DP}} + \beta_2\mathcal{L}_{\text{CTP}} + \beta_3\mathcal{L}_{\text{CTC}}
\label{eq:token_loss}
\end{align}

The token-prediction modules are pre-trained on native-only data and fine-tuned on semi-synthesized parallel data.

\subsection{Sampling Algorithm for Token Conversion}

After processing the source L2-accented tokens with the token encoder, CTP, and DP, we can then initialize the target token sequence and use the DLM decoder to complete it (i.e., ``Target Init.'' and ``DLM Decoder'' in Figure~\ref{fig:overview}). By default, we use the greedy sampler with a threshold-based strategy for reusing source tokens, as demonstrated in Sec.~\ref{sec:threshold_based_reuse} and Algorithm~\ref{alg:dlman_greedy}.

\begin{algorithm}[ht]
\caption{Greedy sampling with CTP-based initialization.}
\label{alg:dlman_greedy}
\small
\begin{algorithmic}[1]
\Require Source tokens $\mathbf{y}^{\text{src}}$ (length $N_{\text{src}}$) and duration ratio $r$
\Statex CTP scores $\{\hat{o}_i\}_{i=1}^{N_{\text{src}}}$, reuse threshold $\tau$, sampling steps $T$
\Statex Content representations $\mathbf{c}$ and CFG strength $w_{\text{DLM}}$
\Ensure Target tokens $\mathbf{y}^{\text{tgt}}$
\State $N_{\text{tgt}} \gets \mathrm{round}(N_{\text{src}}\cdot r)$
\State $K \gets \lceil N_{\text{tgt}}/T \rceil$ \Comment{tokens to unmask per step}
\State $\mathcal{I} \gets \{ i \mid \hat{o}_i > \tau \}$ \Comment{reused source-token indices}
\State Initialize $\mathbf{z}^{(0)}$ by nearest interpolation from source to target length: $\mathbf{z}^{(0)} \in \{1,\dots,V,\texttt{[MASK]}\}^{N_{\text{tgt}}}$
\For{$j=1$ to $N_{\text{tgt}}$}
  \State $i^\star \gets \mathrm{round} \left((j-\tfrac{1}{2})\tfrac{N_{\text{src}}}{N_{\text{tgt}}}+\tfrac{1}{2}\right)$ \Comment{source index}
  \If{$i^\star \in \mathcal{I}$}
    \State $z^{(0)}_j \gets y^{\text{src}}_{i^\star}$
  \Else
    \State $z^{(0)}_j \gets \texttt{[MASK]}$
  \EndIf
\EndFor
\State $N_{\text{mask}} \gets |\{j \mid z^{(0)}_j=\texttt{[MASK]}\}|$
\State $T_{\text{eff}} \gets \left\lceil N_{\text{mask}}/K \right\rceil$, $s_0 \gets \max(1,\, T-T_{\text{eff}}+1)$ \Comment{start step from reuse proportion}
\For{$s=s_0$ to $T$}
  \State Compute logits at masked positions:
  \Statex \hspace{1em}$\boldsymbol{\ell}_{\text{cond}}(\cdot\mid\mathbf{z}^{(s-1)},\mathbf{c})$ and $\boldsymbol{\ell}_{\text{uncond}}(\cdot\mid\mathbf{z}^{(s-1)})$
  \State Apply CFG: $\boldsymbol{\ell}_{\text{cfg}} \gets (1+w_{\text{DLM}})\,\boldsymbol{\ell}_{\text{cond}} - w_{\text{DLM}}\,\boldsymbol{\ell}_{\text{uncond}}$
  \State For each masked position $j$, set $\hat{y}_j \gets \arg\max \boldsymbol{\ell}_{\text{cfg}}(j)$ and confidence $\gamma_j \gets \max\,\mathrm{softmax}(\boldsymbol{\ell}_{\text{cfg}}(j))$
  \State Select $\mathcal{J}$ as the top-$\min(K,\,|\text{masked}|)$ masked positions by $\gamma_j$
  \State Unmask: set $z^{(s)}_j \gets \hat{y}_j$ for $j\in\mathcal{J}$; keep all other positions unchanged
\EndFor
\State \Return $\mathbf{y}^{\text{tgt}} \gets \mathbf{z}^{(T)}$
\end{algorithmic}
\end{algorithm}

\subsection{Token-to-Speech Synthesis}

We use a flow-matching speech synthesizer with a vocoder~\cite{li2023hiftnet} to generate waveforms. The input token sequence is encoded by a relative-positional Transformer encoder, and the encoded features are concatenated with a speaker embedding before being fed into a DiT decoder. The speech synthesizer is trained separately on native-only speech data.
Similar to CosyAccent, DLM-AN deploys a two-way CFG strategy for generation:
\begin{align}
  \bar{v}_\eta (\mathbf{x}_t, t, \mathbf{y}, \mathbf{s}) & = \ v_{\eta}(\mathbf{x}_t, t, \mathbf{y}, \mathbf{s})\nonumber \\
  & + w_1 (v_{\eta}(\mathbf{x}_t, t, \mathbf{y}, \mathbf{s}) - v_{\eta}(\mathbf{x}_t, t, \varnothing, \mathbf{s}))\nonumber \\
  & + w_2 (v_{\eta}(\mathbf{x}_t, t, \mathbf{y}, \mathbf{s}) - v_{\eta}(\mathbf{x}_t, t, \mathbf{y}, \varnothing))
\end{align}
\label{eq:synthesizer_cfg}
where $v_{\eta}$ is the synthesizer, $t \in [0,1]$ is the time variable, $\mathbf{x}_t$ is the Mel-spectrogram at time $t$, $\mathbf{y}$ is the input tokens, and $\mathbf{s}$ is the speaker embedding. The two CFG factors $w_1$ and $w_2$ control the emphasis on the content and timbre conditions, respectively.

\section{Experimental Setup}
\subsection{Datasets}
\label{sec:dataset}
The experiments are conducted on English. Training uses the English subset of Emilia~\cite{he2024emilia} (Emilia-EN) and the LibriTTS-R corpus~\cite{koizumi2023librittsr} with synthesized L2-accented counterparts\footnote{https://huggingface.co/datasets/Piping/L2-LibriTTSR/}~\cite{bai2026cosyaccent}. We also use the L2-ARCTIC corpus~\cite{zhao2018l2arctic} together with four American speakers from ARCTIC~\cite{kominek2004cmu}. We further synthesize pseudo native targets for this extended L2-ARCTIC set, which are used for supervised fine-tuning and evaluation.
The pseudo native targets are generated using a native-only zero-shot Matcha-TTS~\cite{mehta2024matcha} model trained on LibriTTS-R.

Emilia-EN is solely used for pretraining. LibriTTS-R is used to train the SSL tokenizer and the flow-matching speech synthesizer. For fine-tuning, both augmented LibriTTS-R (with synthesized source utterances) and extended L2-ARCTIC (with synthesized targets) are utilized.

Since the L2-accented counterparts~\cite{bai2026cosyaccent} of LibriTTS-R were synthesized using prompts drawn from L2-ARCTIC, we adopt the same train-valid-test partition of L2-ARCTIC to prevent text leakage.

\subsection{Tokenizer}
The proposed method relies on the phonetic richness of the speech tokens. We use WavLM$_\text{large}$ and extract layer-22 representations for tokenization. We train an online K-Means model with 1024 clusters (codebook size 1024) on LibriTTS-R.

\begin{figure*}[htbp]
\centering
\includegraphics[width=0.99\textwidth]{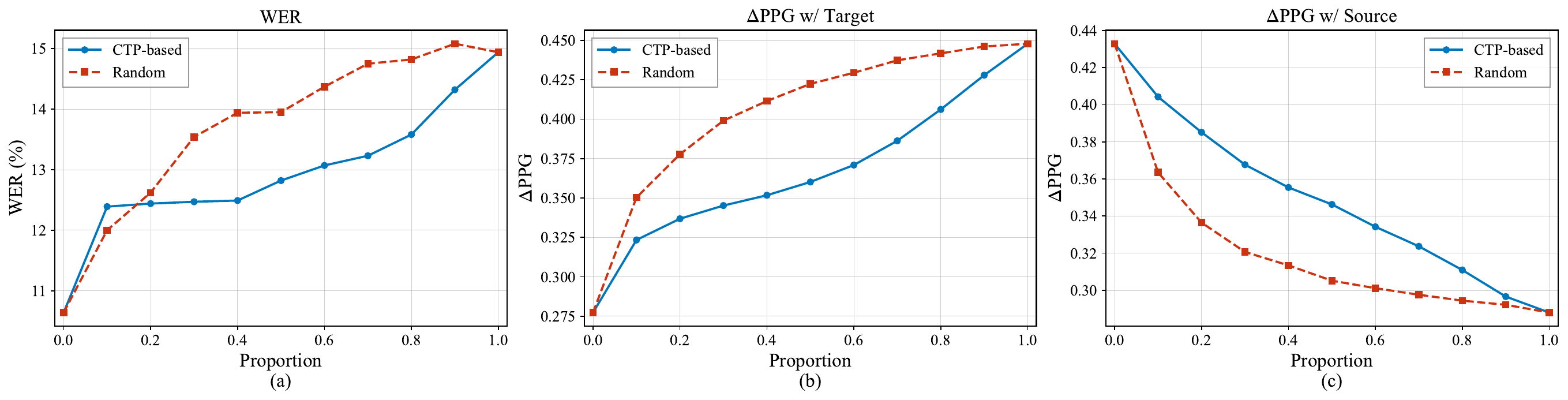}
\vspace{-0.3cm}
\caption{CTP-based vs.\ random token selection at varying reuse proportions. Three metrics are compared: (a) WER, (b) $\Delta$PPG with the L1-accented target, and (c) $\Delta$PPG with the L2-accented source. CTP-based selection achieves generally lower WER and consistently better accent separation than random selection at the same proportion.}
\label{fig:ctp_vs_random}
\vspace{-0.5cm}
\end{figure*}

\subsection{Compared Systems}
We evaluate our model against two strong baselines:
\begin{itemize}
  \item \textit{TokAN}~\cite{bai2025accent}: An autoregressive model operating on deduplicated tokens. It recovers token-wise durations in the speech synthesis stage. This model features a flow matching duration predictor with two conditions: 1) the token sequence and 2) the average token duration. When provided with the average token duration, TokAN is able to preserve the total duration.
  We test two modes: \textit{\textbf{TokAN-1}}, which predicts token durations directly, and \textit{\textbf{TokAN-2}}, which predicts with total-duration awareness and preserves the total duration.
  \item \textit{CosyAccent}~\cite{bai2026cosyaccent}: A non-autoregressive direct flow-matching model. It features a total-duration ratio predictor similar to our proposed model. It can be regarded as a continuous-diffusion counterpart of the proposed model, but without accent-strength control. We test two modes: \textit{\textbf{CosyAccent-1}}, which predicts the total duration ratio, and \textit{\textbf{CosyAccent-2}}, which inherits the source total duration.
\end{itemize}

% Both baseline systems were trained on paired data consisting of real L2 speech (from L2-ARCTIC and the Chinese-accented set) and their corresponding L1 targets from LibriTTS-R.

For the token-reuse setting of DLM-AN, we deploy a threshold-based selection strategy: tokens with CTP confidence higher than a threshold ($\tau$) are reused. For simplicity, we do not apply token reuse and duration scaling simultaneously. We evaluate four configurations of DLM-AN:
\begin{itemize}
    \item \textbf{\textit{DLM-AN-1}}: Uses the predicted total duration ratio.
    \item \textbf{\textit{DLM-AN-2} ($\tau=1.0$)}: Inherits the source total duration while predicting the target sequence from scratch.
    \item \textbf{\textit{DLM-AN-2 ($\tau=0.3$)}}: Inherits the source total duration, with a threshold for common token prediction (i.e., $\tau=0.3$).
    \item \textbf{\textit{DLM-AN-2 ($\tau=0.0$)}}: Inherits all the source tokens (i.e., direct resynthesis).
\end{itemize}

For TokAN and DLM-AN, the two token-based systems, we pretrain them on Emilia-EN.
BART-style~\cite{lewis2019bart,jia2024convert} token corruption is applied to the source token sequence. This helps avoid teaching the model to trivially copy the source sequence. Different from the original TokAN, no source accent embedding is utilized. Given the corrupted source sequence, TokAN performs autoregressive generation while DLM-AN performs masked generation. TokAN and DLM-AN share a similar architecture: a speech token encoder with CTC-based phonemic guidance, a self-attention-only decoder, and a flow-matching speech synthesizer. TokAN predicts token-wise durations in the synthesizer module, whereas DLM-AN predicts the total duration ratio during token-level conversion.
For DLM-AN, we set the loss weights in Eq.\,(\ref{eq:token_loss}) as \(\beta_1{=}1.0\), \(\beta_2{=}1.0\), and \(\beta_3{=}0.2\) for joint training of the token-prediction modules. For CTP training, we use a positive weight of 2 for better label balance.

For TokAN generation, we use beam search with a beam size of 10. For DLM-AN, we use the greedy sampler with 32 steps; the CFG guidance strength $w_\text{DLM}$ is set to 1.0. For the speech synthesizers in TokAN and DLM-AN, we use the Euler sampler with 32 steps. The CFG strengths $w_1$ and $w_2$ are set to 1.0 and 1.0, respectively.

For CosyAccent, we use the official Whisper$_\text{medium}$ model~\cite{radford2023robust} as the frozen speech frontend. CosyAccent is also pre-trained on Emilia-EN, with the encoder supervised only by the CTC loss. During inference, we use the same CFG weights and number of sampling steps as in the original paper.

Resemblyzer\footnote{https://github.com/resemble-ai/Resemblyzer} is deployed to extract speaker embeddings for the speech synthesizer modules in all the compared models. The final waveform is generated using the HiFTNet vocoder~\cite{li2023hiftnet} from CosyVoice2~\cite{du2024cosyvoice}.

\subsection{Evaluation Data \& Metrics}

\noindent\textbf{Evaluation Set.}
Our test set involves the extended L2-ARCTIC dataset, as described in Sec.~\ref{sec:dataset}. The test set covers seven accents: Arabic, Chinese, Hindi, Korean, Spanish, Vietnamese, and native American English. The set contains 80 sentences. The partitioning is consistent with the source-synthesis training data~\cite{bai2026cosyaccent}, with no text leakage in the training data.

\noindent\textbf{Subjective Evaluation.}
We conducted listening tests with 25 raters to assess three qualities: \textit{Naturalness (NAT)} and \textit{Accentedness (ACT)} were measured via MUSHRA tests. The native accent was excluded from the ACT evaluation. \textit{Speaker Similarity (SIM)} was measured via Best-Worst Scaling (BWS), with scores aggregated using a standard counting algorithm~\cite{ravillion2020comparison}: $(N_\textit{best} - N_\textit{worst}) / N_\textit{occurrence}$.

\noindent\textbf{Objective Evaluation.}
We use four objective metrics to assess conversion quality automatically. \textit{Intelligibility:} Word Error Rate (WER) from a native-only ASR model\footnote{https://huggingface.co/facebook/s2t-medium-librispeech-asr} to simulate listener perception. \textit{Naturalness:} The UTMOSv2 score\footnote{https://github.com/sarulab-speech/UTMOSv2} from a neural naturalness predictor. \textit{Timbre Preservation:} Speaker Encoding Cosine Similarity (SECS) using the accent-robust Resemblyzer. \textit{Accentedness Reduction:} The phonetic posteriorgram distance ($\Delta$PPG)\footnote{https://github.com/interactiveaudiolab/ppgs}~\cite{churchwell2024high}. By default, $\Delta$PPG is computed between generated utterances and the synthesized native targets, but some analyses also compute it against the source to measure how much accent is removed.

\begin{table*}[htb]
  \centering
  \caption{Evaluation results of accent normalization systems. Source-length indicates whether the source total duration is preserved. $\tau$ denotes the CTP threshold for token reuse ($\tau{=}1.0$: generation from scratch; $\tau{=}0.0$: full token reuse / resynthesis). Best and second-best objective results are in \textbf{bold} and \underline{underlined}.}
  \vspace{-0.2cm}
  \small
  \begin{tabular}{lcccccccc}
    \toprule
    \multirow{2}{*}{System} & \multirow{2}{*}{Source-length} & \multicolumn{3}{c}{Subjective}&\multicolumn{4}{c}{Objective} \\
    \cmidrule(r){3-5}\cmidrule(r){6-9}
    & & NAT ($\uparrow$) & ACT ($\downarrow$) & SIM ($\uparrow$) & WER (\% $\downarrow$) & UTMOS ($\uparrow$) & SECS ($\uparrow$) & $\Delta\text{PPG}$ ($\downarrow$) \\
    \midrule
    Source & $\checkmark$ & 58.36{\tiny$\pm$2.84} & 48.46{\tiny$\pm$2.61} & - & 15.86 & 2.80{\tiny$\pm$0.41} & - & 0.5097 \\
    \midrule
    TokAN-1~\cite{bai2025accent} & $\times$ & {\bf 63.85}{\tiny$\pm$2.49} & \underline{23.58}{\tiny$\pm$1.85} & -0.082 & 13.82 & {\bf 3.07}{\tiny$\pm$0.36} & 0.8495 & 0.2884 \\
    TokAN-2~\cite{bai2025accent} & $\checkmark$ & 59.20{\tiny$\pm$2.60} & 28.01{\tiny$\pm$2.21} & -0.051 & 14.00 & 2.97{\tiny$\pm$0.38} & 0.8530 & 0.2980 \\
    \midrule
    CosyAccent-1~\cite{bai2026cosyaccent} & $\times$ & 61.12{\tiny$\pm$2.42} & 25.75{\tiny$\pm$1.89} & -0.071 & 12.40 & 2.99{\tiny$\pm$0.34} & 0.8294 & {\bf 0.2736} \\
    CosyAccent-2~\cite{bai2026cosyaccent} & $\checkmark$ & 56.35{\tiny$\pm$2.61} & 29.51{\tiny$\pm$2.12} & \phantom{-}0.051 & 13.84 & 2.89{\tiny$\pm$0.36} & 0.8358 & 0.3029 \\
    \midrule
    DLM-AN-1 & $\times$ & \underline{62.20}{\tiny$\pm$2.46} & {\bf 22.94}{\tiny$\pm$1.85} & -0.133 & \underline{11.19} & \underline{3.05}{\tiny$\pm$0.34} & 0.8385 & 0.2811 \\
    DLM-AN-2 ($\tau$=1.0) & $\checkmark$ & 59.50{\tiny$\pm$2.61} & 27.90{\tiny$\pm$2.16} & -0.020 &  {\bf 10.64} & 2.93{\tiny$\pm$0.37} & 0.8521 & \underline{0.2773} \\
    DLM-AN-2 ($\tau$=0.3) & $\checkmark$ & 57.35{\tiny$\pm$2.59} & 31.34{\tiny$\pm$2.34} & \phantom{-}\underline{0.098} & 12.52 & 2.90{\tiny$\pm$0.39} & \underline{0.8590} & 0.3580 \\
    DLM-AN-2 ($\tau$=0.0) & $\checkmark$ & 56.41{\tiny$\pm$2.70} & 38.37{\tiny$\pm$2.57} & \phantom{-}{\bf 0.208} & 14.94 & 2.86{\tiny$\pm$0.39} & {\bf 0.8646} & 0.4479 \\
    \bottomrule
  \end{tabular}
\label{tab:res}
\vspace{-0.2cm}
\end{table*}

\section{Results}

\subsection{Effectiveness of Common Token Prediction}

For effective common token prediction, higher confidence should be assigned to native-accented regions, whereas low confidence scores should be assigned to highly-L2-accented regions. Figure~\ref{fig:lcs_vis} is a visualization of common token prediction for a Chinese-accented sample. The common token confidence values are displayed over the spectrogram. Phonemes predicted by ppgs~\cite{churchwell2024high} are displayed below the spectrogram, with their boundaries being the white dashed lines on the Mel-spectrogram. Aligned words, obtained via MFA~\cite{mcauliffe2017montreal}, are displayed at the bottom. Accent can be determined from the comparison between the words and phonemes. Some prominent L2-accented patterns can be found: 1) the initial word ``a" is heavily lengthened; correspondingly, the CTP confidence becomes low in the prolonged part. 2) The PPG-predicted phonemes for the word ``had" is messy, corresponding with general low confidence scores. 3) The ending phoneme of ``death" is detected as highly similar to /S/, receiving low confidence scores.

\subsubsection{Proportion-based Reuse}

To progressively preserve or remove the source accent, we can directly control the proportion of reused source tokens based on the CTP confidence scores. However, random reuse may also yield a coarse accent-retention effect. To verify the benefit of CTP-based reuse, we vary the reuse proportion under two strategies---CTP-based selection and random selection. Figure~\ref{fig:ctp_vs_random} reports three metrics: 1) WER, 2) $\Delta$PPG to the L1-accented target, and 3) $\Delta$PPG to the L2-accented source. Ideally, WER should increase as the reuse proportion increases. In contrast, $\Delta$PPG to the target should decrease (more native-like), while $\Delta$PPG to the source should increase (less similar to the accented input).

Both strategies follow these general trends, but their outcomes differ. CTP-based selection generally yields lower WER than random selection at the same reuse proportion. Moreover, it consistently achieves higher $\Delta$PPG to the source and lower $\Delta$PPG to the target, indicating better accent removal while remaining closer to the native reference. This suggests that CTP indeed prioritizes more native-accented regions for reuse, whereas random selection more frequently preserves highly accented tokens. Overall, CTP-based token reuse provides a more reliable control knob for progressive accent normalization.

\subsubsection{Threshold-based Reuse}
\label{sec:threshold_based_reuse}

Input utterances exhibit different strengths of L2 accent. Therefore, enforcing the same reuse proportion for all sources can be suboptimal. We instead adopt threshold-based reuse, where tokens are reused if their CTP confidence exceeds a threshold $\tau$ (Algorithm~\ref{alg:dlman_greedy}). This makes the effective reuse proportion adaptive to the input accent strength.

\begin{figure}[htbp]
\centering
\vspace{-0.2cm}
\includegraphics[width=0.38\textwidth]{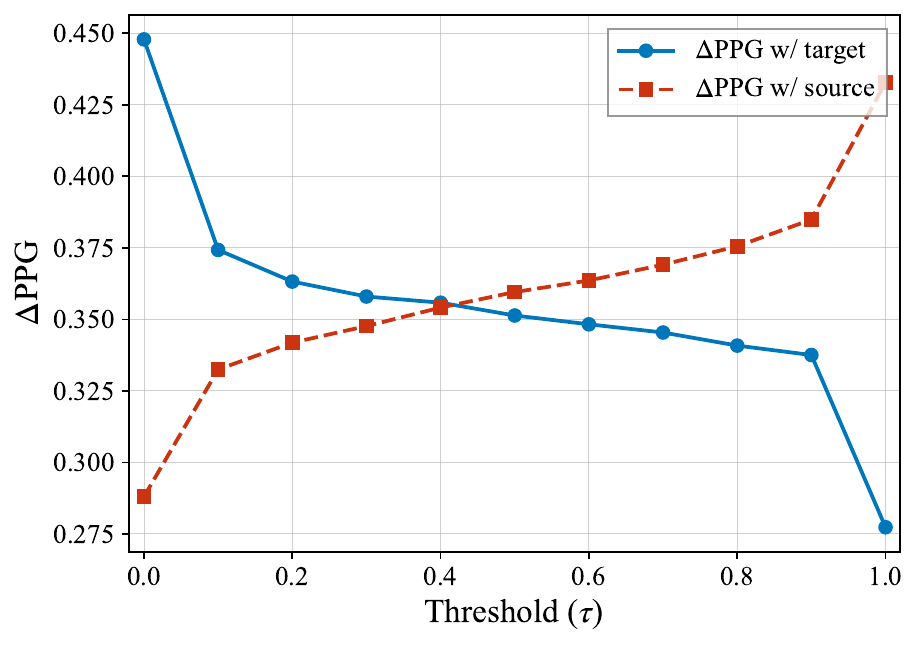}
\vspace{-0.3cm}
\caption{$\Delta$PPG with the L2-accented source and L1-accented target at varying CTP thresholds $\tau$. Lower $\tau$ retains more source tokens. As $\tau$ increases, $\Delta$PPG with the source increases (more accent removed) while $\Delta$PPG with the target decreases (closer to native).}
\vspace{-0.3cm}
\label{fig:threshold_ppg}
\end{figure}

Figure~\ref{fig:threshold_ppg} shows $\Delta$PPG to the source and to the L1-accented target under different thresholds. When $\tau{=}0.0$, all tokens are reused; as $\tau$ increases, fewer tokens are kept; and when $\tau{=}1.0$, generation from scratch is performed. As expected, $\Delta$PPG to the source increases monotonically with $\tau$, while $\Delta$PPG to the target decreases, confirming that $\tau$ provides an interpretable and effective knob for accent reduction.

% \subsubsection{Average Confidence over Accents}

\subsection{Comparison with Baselines}

\subsubsection{Duration-free Conversion}

The main results are shown in Table~\ref{tab:res}. We first compare the systems under the \textit{free duration} setting (``-1'' variants), where the model predicts its own target duration. DLM-AN-1 achieves the lowest ACT score (22.94) among all systems, indicating the strongest accent reduction, while attaining the second-highest NAT score (62.20), close to TokAN-1 (63.85) and above CosyAccent-1 (61.12). In terms of objective content preservation, DLM-AN-1 obtains a WER of 11.19\%, notably lower than TokAN-1 (13.82\%) and CosyAccent-1 (12.40\%). The UTMOS score (3.05) is also competitive with TokAN-1 (3.07), both outperforming CosyAccent-1 (2.99). These results suggest that DLM-AN achieves the best overall balance of accent reduction and content preservation under free duration.

\subsubsection{Source-duration-preserved Conversion}

Under the \textit{source-duration-preserved} setting (``-2'' variants), DLM-AN-2 ($\tau{=}1.0$) again achieves the best WER (10.64\%) and a competitive $\Delta$PPG (0.2773) among all systems. As expected, preserving the source duration generally leads to higher ACT scores and lower NAT scores compared with free-duration counterparts across all models, as the L2-accented rhythm is retained. Notably, DLM-AN-2 ($\tau{=}1.0$) attains a similar $\Delta$PPG with DLM-AN-1 (0.2773 vs.\ 0.2811) yet a higher ACT (27.90 vs.\ 22.94). This suggests that $\Delta$PPG primarily captures segmental (more phonetic) similarity, whereas human raters perceive accentedness more holistically---rhythm and prosody, which are largely determined by duration, play a substantial role in subjective accent judgments.

\subsubsection{Controllable Accent Normalization}

The effect of the CTP-based token reuse is clearly visible across the DLM-AN-2 variants. As the threshold $\tau$ decreases from 1.0 to 0.0, more source tokens are preserved, leading to a progressive increase in ACT (27.90 $\to$ 31.34 $\to$ 38.37), reflecting stronger retention of the source accent. Correspondingly, the SIM score rises monotonically ($-0.020 \to 0.098 \to 0.208$), confirming that preserving more source tokens improves perceived speaker similarity. The SECS scores follow the same trend (0.8521 $\to$ 0.8590 $\to$ 0.8646), with the full-reuse variant achieving the highest timbre preservation. This correlation between the source accent and speaker identity aligns with the finding in \cite{halychanskyi2025fac}.

Meanwhile, WER degrades mildly (10.64 $\to$ 12.52 $\to$ 14.94) and $\Delta$PPG increases (0.2773 $\to$ 0.3580 $\to$ 0.4479), as reusing accented tokens inevitably retains some non-native pronunciation patterns. At $\tau{=}0.0$ (complete resynthesis), the output closely mirrors the source accent, as indicated by the ACT score (38.37) approaching the source (48.46), demonstrating smooth and interpretable accent strength control.

\subsection{Arbitrary Duration Scaling}

All three systems support total-duration specification, but via different mechanisms: TokAN predicts token-wise durations after accent normalization, CosyAccent directly generates a target-length spectrogram, and DLM-AN generates a target-length token sequence. To compare robustness under different target lengths, we vary the total-duration ratio and compare WER across systems.

\begin{figure}[htbp]
\centering
\includegraphics[width=0.38\textwidth]{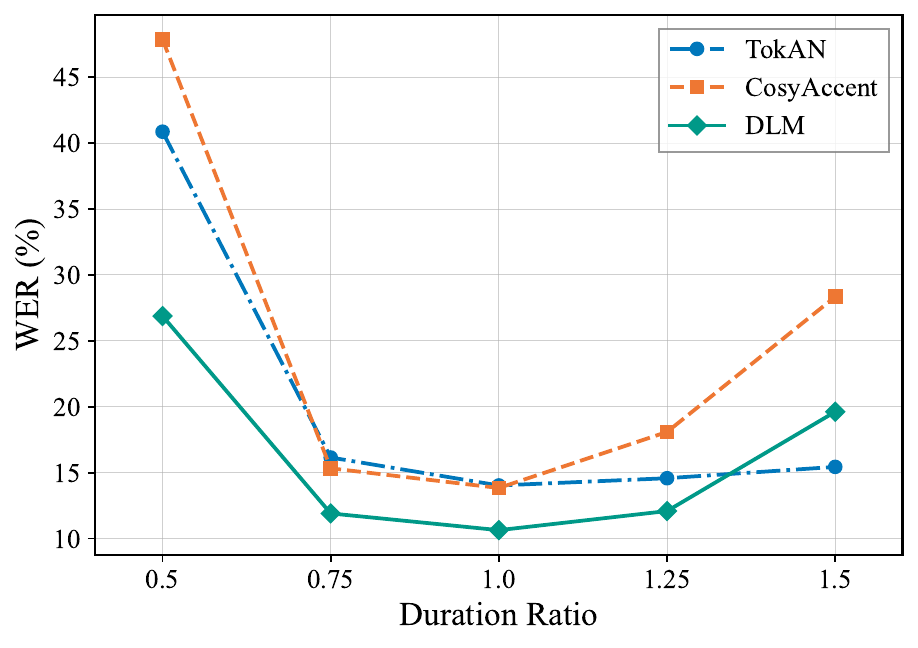}
\vspace{-0.3cm}
\caption{WER (\%) at varying duration scaling ratios for TokAN, CosyAccent, and DLM-AN. DLM-AN achieves the lowest WER across most ratios, with a notable advantage under compression (ratio $< 1.0$).}
% \vspace{-0.5cm}
\label{fig:duration_scaling_wer}
\end{figure}

Figure~\ref{fig:duration_scaling_wer} shows the results. DLM-AN achieves the lowest WER (i.e., best content preservation) when the source duration is preserved, and its advantage is more pronounced when the specified ratio is smaller than \SI{1.0}{}. When the target duration is set to half of the source, TokAN degrades substantially because the generated token sequence (after deduplication) is often longer than the desired duration, forcing tokens to be discarded in the synthesis stage. DLM-AN maintains an advantage until the ratio reaches \SI{1.5}{}, likely because such extreme stretching is rare in the training data.

% \subsection{Corrective Sampling for Few-step Conversion}

\subsection{Ablation Study}

The ablation results are shown in Table~\ref{tab:ablation_res}, which is based on DLM-AN-2 that generates target tokens from scratch. We ablate two factors: 1) CFG in token generation, 2) pretraining on Emilia-EN.
The results show the effectiveness of these training/inference components in the proposed DLM-AN.

\begin{table}[!htb]
  \centering
  \caption{Ablation results of DLM-AN-2 ($\tau$=1.0).}
  \label{tab:ablation_res}
  \setlength{\tabcolsep}{4pt}  % Reduce column spacing (default is 6pt)
  \begin{tabular}{lccc}
    \toprule
    System & WER ($\% \downarrow$) & SECS ($\uparrow$) & $\Delta\text{PPG}$ ($\downarrow$) \\
    \midrule
    Source & 15.86 & - & 0.5097 \\
    \midrule
    DLM-AN & {\bf 10.64} & {\bf 0.8521} & {\bf 0.2773} \\
    \quad w/o CFG & 11.23 & 0.8488 & 0.2907 \\
    \quad w/o pretraining & 16.61 & 0.8513 & 0.3306 \\
    \bottomrule
\end{tabular}
\end{table}

% \section{Analysis}

% \begin{itemize}
%     \item Token distribution - source
%     \item Token distribution - thr-1.0  (no reuse)
%     \item Token distribution - thr-0.3
%     \item Token distribution - thr-0.0  (resynthesis)
% \end{itemize}

\section{Conclusion \& Future Work}
We presented DLM-AN, a controllable accent normalization system based on masked discrete diffusion over self-supervised speech tokens. By introducing a Common Token Predictor (CTP) that identifies source tokens likely shared with the native target, DLM-AN provides a simple yet effective accent-strength knob: reusing more high-confidence tokens preserves more of the original accent, while generating all tokens from scratch yields full normalization. A duration ratio predictor further enables total-duration adjustment. Experiments on multi-accent English data show that DLM-AN achieves the lowest WER among all compared systems, competitive naturalness and accent reduction, and smooth, interpretable accent strength control across a continuous range.

Several directions remain for future work. First, the current pipeline relies on a recognition-based token encoder for phoneme supervision, whose errors can propagate and degrade conversion quality for heavily accented inputs.
Second, repeated pronunciations can arise during unmasking; incorporating corrective mechanisms~\cite{huang2025dont,bie2026llada2} may mitigate such artifacts.
Third, the SSL tokenizer and synthesizer are trained on native-only data, potentially limiting reconstruction of highly accented speech; incorporating L2-accented data could improve robustness. Forth, replacing the K-Means tokenizer with learned discrete codebooks (e.g., vector quantization~\cite{van2017vq}) may yield better phonetic discriminability and further enhance controllability.

\vfill
\pagebreak

\section{Acknowledgments}
This research is supported by National Natural Science Foundation of China (Grant No. 62401377 and No. 62271432), Program for Guangdong Introducing Innovative and Entrepreneurial Teams (Grant No. 2023ZT10X044), Yangtze River Delta Science and Technology Innovation Community Joint Research Project (Grant No. 2024CSJGG1100), Shenzhen Science and Technology Program (Shenzhen Key Laboratory, Grant No. ZDSYS20230626091302006), Shenzhen Stability Science Program 2023, Shenzhen Key Lab of Multi-Modal Cognitive Computing, and the internal project of the Guangdong Provincial Key Laboratory of Big Data Computing (Grant No. B10120210117-KP02), The Chinese University of Hong Kong, Shenzhen (CUHK-Shenzhen).

\section{Generative AI Use Disclosure}
Generative AI tools were used solely for editing and polishing the manuscript text. No part of the scientific content, including the ideas, methodology, experiments, or analysis, was generated by AI. All authors have reviewed and take full responsibility for the content of this paper.

\bibliographystyle{IEEEtran}
\bibliography{mybib}

@article{hsu2021hubert,
  title     = {Hubert: Self-supervised speech representation learning by masked prediction of hidden units},
  author    = {Hsu, Wei-Ning and Bolte, Benjamin and Tsai, Yao-Hung Hubert and Lakhotia, Kushal and Salakhutdinov, Ruslan and Mohamed, Abdelrahman},
  journal   = {TASLP},
  volume    = {29},
  pages     = {3451--3460},
  year      = {2021},
  publisher = {IEEE}
}

@article{chen2022wavlm,
  title     = {Wavlm: Large-scale self-supervised pre-training for full stack speech processing},
  author    = {Chen, Sanyuan and Wang, Chengyi and Chen, Zhengyang and Wu, Yu and Liu, Shujie and Chen, Zhuo and Li, Jinyu and Kanda, Naoyuki and Yoshioka, Takuya and Xiao, Xiong and others},
  journal   = {J-STSP},
  volume    = {16},
  number    = {6},
  pages     = {1505--1518},
  year      = {2022},
  publisher = {IEEE}
}

@article{lewis2019bart,
  title   = {Bart: Denoising sequence-to-sequence pre-training for natural language generation, translation, and comprehension},
  author  = {Lewis, M},
  journal = {arXiv preprint arXiv:1910.13461},
  year    = {2019}
}

@inproceedings{choi24self,
  title     = {Self-Supervised Speech Representations are More Phonetic than Semantic},
  author    = {Kwanghee Choi and Ankita Pasad and Tomohiko Nakamura and Satoru Fukayama and Karen Livescu and Shinji Watanabe},
  year      = {2024},
  booktitle = {Proc. Interspeech}
}

@article{lakhotia2021generative,
  title   = {On generative spoken language modeling from raw audio},
  author  = {Lakhotia, Kushal and Kharitonov, Eugene and Hsu, Wei-Ning and Adi, Yossi and Polyak, Adam and Bolte, Benjamin and Nguyen, Tu-Anh and Copet, Jade and Baevski, Alexei and Mohamed, Abdelrahman and others},
  journal = {Trans. ACL},
  volume  = {9},
  pages   = {1336--1354},
  year    = {2021}
}

@inproceedings{radford2023robust,
  title     = {Robust speech recognition via large-scale weak supervision},
  author    = {Radford, Alec and Kim, Jong Wook and Xu, Tao and Brockman, Greg and McLeavey, Christine and Sutskever, Ilya},
  booktitle = {ICML},
  pages     = {28492--28518},
  year      = {2023}
}

@inproceedings{graves2006connectionist,
  title     = {Connectionist temporal classification: labelling unsegmented sequence data with recurrent neural networks},
  author    = {Graves, Alex and Fern{\'a}ndez, Santiago and Gomez, Faustino and Schmidhuber, J{\"u}rgen},
  booktitle = {ICML},
  year      = {2006}
}

@inproceedings{cui2025exploring1,
  title     = {Exploring SSL discrete tokens for multilingual ASR},
  author    = {Cui, Mingyu and Tan, Daxin and Yang, Yifan and Wang, Dingdong and Wang, Huimeng and Chen, Xiao and Chen, Xie and Liu, Xunying},
  booktitle = {Proc. ICASSP 2025},
  year      = {2025}
}

@inproceedings{cui2025exploring2,
  title     = {Exploring ssl discrete speech features for zipformer-based contextual asr},
  author    = {Cui, Mingyu and Yang, Yifan and Deng, Jiajun and Kang, Jiawen and Hu, Shujie and Wang, Tianzi and Li, Zhaoqing and Zhang, Shiliang and Chen, Xie and Liu, Xunying},
  booktitle = {Proc. Interspeech 2025},
  year      = {2025}
}

@inproceedings{onda2026advanced,
  title     = {Advanced Modeling of Interlanguage Speech Intelligibility Benefit with L1-L2 Multi-Task Learning Using Differentiable K-Means for Accent-Robust Discrete Token-Based ASR},
  author    = {Onda, Kentaro and Fukayama, Satoru and Saito, Daisuke and Minematsu, Nobuaki},
  booktitle = {Proc. ICASSP 2026},
  year      = {2026}
}

@inproceedings{huang2021any,
  title     = {Any-to-one sequence-to-sequence voice conversion using self-supervised discrete speech representations},
  author    = {Huang, Wen-Chin and Wu, Yi-Chiao and Hayashi, Tomoki},
  booktitle = {Proc. ICASSP},
  year      = {2021}
}

@inproceedings{kreuk2022textless,
  title     = {Textless Speech Emotion Conversion using Discrete \& Decomposed Representations},
  author    = {Kreuk, Felix and Polyak, Adam and Copet, Jade and Kharitonov, Eugene and Nguyen, Tu-Anh and Rivi{\`e}re, Morgan and Hsu, Wei-Ning and Mohamed, Abdelrahman and Dupoux, Emmanuel and Adi, Yossi},
  booktitle = {Proc. EMNLP},
  year      = {2022}
}

@article{oh2025durflex,
  title     = {DurFlex-EVC: Duration-Flexible Emotional Voice Conversion Leveraging Discrete Representations Without Text Alignment},
  author    = {Oh, Hyung-Seok and Lee, Sang-Hoon and Cho, Deok-Hyeon and Lee, Seong-Whan},
  journal   = {IEEE Transactions on Affective Computing},
  year      = {2025},
  publisher = {IEEE}
}

@article{li2023hiftnet,
  title   = {Hiftnet: A fast high-quality neural vocoder with harmonic-plus-noise filter and inverse short time fourier transform},
  author  = {Li, Yinghao Aaron and Han, Cong and Jiang, Xilin and Mesgarani, Nima},
  journal = {arXiv preprint arXiv:2309.09493},
  year    = {2023}
}

@inproceedings{mehta2024matcha,
  title     = {Matcha-TTS: A fast TTS architecture with conditional flow matching},
  author    = {Mehta, Shivam and Tu, Ruibo and Beskow, Jonas and Sz{\'e}kely, {\'E}va and Henter, Gustav Eje},
  booktitle = {Proc. ICASSP},
  pages     = {11341--11345},
  year      = {2024}
}

@article{kharitonov2023speak,
  title   = {Speak, read and prompt: High-fidelity text-to-speech with minimal supervision},
  author  = {Kharitonov, Eugene and Vincent, Damien and Borsos, Zal{\'a}n and Marinier, Rapha{\"e}l and Girgin, Sertan and Pietquin, Olivier and Sharifi, Matt and Tagliasacchi, Marco and Zeghidour, Neil},
  journal = {Trans. ACL},
  volume  = {11},
  pages   = {1703--1718},
  year    = {2023}
}

@inproceedings{chang2022maskgit,
  title     = {Maskgit: Masked generative image transformer},
  author    = {Chang, Huiwen and Zhang, Han and Jiang, Lu and Liu, Ce and Freeman, William T},
  booktitle = {Proceedings of the IEEE/CVF conference on computer vision and pattern recognition},
  pages     = {11315--11325},
  year      = {2022}
}

@inproceedings{wang2025maskgct,
  title     = {Mask{GCT}: Zero-Shot Text-to-Speech with Masked Generative Codec Transformer},
  author    = {Yuancheng Wang and Haoyue Zhan and Liwei Liu and Ruihong Zeng and Haotian Guo and Jiachen Zheng and Qiang Zhang and Xueyao Zhang and Shunsi Zhang and Zhizheng Wu},
  booktitle = {ICLR},
  year      = {2025}
}

@inproceedings{wang2025metis,
  title     = {Metis: A Foundation Speech Generation Model with Masked Generative Pre-training},
  author    = {Yuancheng Wang and Jiachen Zheng and Junan Zhang and Xueyao Zhang and Huan Liao and Zhizheng Wu},
  booktitle = {The Thirty-ninth Annual Conference on Neural Information Processing Systems},
  year      = {2025}
}

@article{fang2024llamaomni,
  title   = {LLaMA-Omni: Seamless Speech Interaction with Large Language Models},
  author  = {Qingkai Fang and Shoutao Guo and Yan Zhou and Zhengrui Ma and Shaolei Zhang and Yang Feng},
  year    = {2024},
  journal = {arXiv preprint arXiv:2409.06666}
}

@inproceedings{hoogeboom2021argmax,
  title     = {Argmax flows and multinomial diffusion: Learning categorical distributions},
  author    = {Hoogeboom, Emiel and Nielsen, Didrik and Jaini, Priyank and Forr{\'e}, Patrick and Welling, Max},
  booktitle = {Advances in neural information processing systems},
  volume    = {34},
  pages     = {12454--12465},
  year      = {2021}
}

@inproceedings{benhamu2025accelerated,
  title     = {Accelerated Sampling from Masked Diffusion Models via Entropy Bounded Unmasking},
  author    = {Heli Ben-Hamu and Itai Gat and Daniel Severo and Niklas Nolte and Brian Karrer},
  booktitle = {The Thirty-ninth Annual Conference on Neural Information Processing Systems},
  year      = {2025},
  url       = {https://openreview.net/forum?id=WBcBhT1NKO}
}

@article{huang2025dont,
  title   = {Don't Settle Too Early: Self-Reflective Remasking for Diffusion Language Models},
  author  = {Huang, Zemin and Wang, Yuhang and Chen, Zhiyang and Qi, Guo-Jun},
  journal = {arXiv preprint arXiv:2509.23653},
  year    = {2025}
}

@article{sahoo2024simple,
  title   = {Simple and effective masked diffusion language models},
  author  = {Sahoo, Subham S and Arriola, Marianne and Schiff, Yair and Gokaslan, Aaron and Marroquin, Edgar and Chiu, Justin T and Rush, Alexander and Kuleshov, Volodymyr},
  journal = {Advances in Neural Information Processing Systems},
  volume  = {37},
  pages   = {130136--130184},
  year    = {2024}
}

@inproceedings{austin2021structured,
  title     = {Structured Denoising Diffusion Models in Discrete State-Spaces},
  author    = {Jacob Austin and Daniel D. Johnson and Jonathan Ho and Daniel Tarlow and Rianne van den Berg},
  booktitle = {Advances in Neural Information Processing Systems},
  year      = {2021}
}

@inproceedings{lou2024discrete,
  title     = {Discrete diffusion modeling by estimating the ratios of the data distribution},
  author    = {Lou, Aaron and Meng, Chenlin and Ermon, Stefano},
  booktitle = {ICML 2024},
  year      = {2024}
}

@article{nie2025llada,
  title   = {Large language diffusion models},
  author  = {Nie, Shen and Zhu, Fengqi and You, Zebin and Zhang, Xiaolu and Ou, Jingyang and Hu, Jun and Zhou, Jun and Lin, Yankai and Wen, Ji-Rong and Li, Chongxuan},
  journal = {arXiv preprint arXiv:2502.09992},
  year    = {2025}
}

@inproceedings{wang2025remasking,
  title     = {Remasking Discrete Diffusion Models with Inference-Time Scaling},
  author    = {Guanghan Wang and Yair Schiff and Subham Sekhar Sahoo and Volodymyr Kuleshov},
  booktitle = {The Thirty-ninth Annual Conference on Neural Information Processing Systems},
  year      = {2025},
  url       = {https://openreview.net/forum?id=IJryQAOy0p}
}

@misc{yu2025discrete,
  title         = {Discrete Diffusion in Large Language and Multimodal Models: A Survey},
  author        = {Runpeng Yu and Qi Li and Xinchao Wang},
  year          = {2025},
  eprint        = {2506.13759},
  archiveprefix = {arXiv},
  primaryclass  = {cs.LG},
  url           = {https://arxiv.org/abs/2506.13759}
}

@article{song2025seed,
  title   = {Seed diffusion: A large-scale diffusion language model with high-speed inference},
  author  = {Song, Yuxuan and Zhang, Zheng and Luo, Cheng and Gao, Pengyang and Xia, Fan and Luo, Hao and Li, Zheng and Yang, Yuehang and Yu, Hongli and Qu, Xingwei and others},
  journal = {arXiv preprint arXiv:2508.02193},
  year    = {2025}
}

@misc{zhang2026corrective,
  title         = {Corrective Diffusion Language Models},
  author        = {Shuibai Zhang and Fred Zhangzhi Peng and Yiheng Zhang and Jin Pan and Grigorios G. Chrysos},
  year          = {2026},
  eprint        = {2512.15596},
  archiveprefix = {arXiv},
  primaryclass  = {cs.LG},
  url           = {https://arxiv.org/abs/2512.15596}
}

@article{bie2026llada2,
  title   = {LLaDA2.1: Speeding Up Text Diffusion via Token Editing},
  author  = {Bie, Tiwei and Cao, Maosong and Cao, Xiang and Chen, Bingsen and Chen, Fuyuan and Chen, Kun and Du, Lun and Feng, Daozhuo and Feng, Haibo and Gong, Mingliang and others},
  journal = {arXiv preprint arXiv:2602.08676},
  year    = {2026}
}

@inproceedings{koizumi2023librittsr,
  author    = {Yuma Koizumi and Heiga Zen and Shigeki Karita and Yifan Ding and Kohei Yatabe and Nobuyuki Morioka and Michiel Bacchiani and Yu Zhang and Wei Han and Ankur Bapna},
  title     = {{LibriTTS-R: A Restored Multi-Speaker Text-to-Speech Corpus}},
  year      = 2023,
  booktitle = {Proc. Interspeech}
}

@inproceedings{zhao2018l2arctic,
  author    = {Guanlong Zhao and Sinem Sonsaat and Alif Silpachai and Ivana Lucic and Evgeny Chukharev-Hudilainen and John Levis and Ricardo Gutierrez-Osuna},
  title     = {{L2-ARCTIC: A Non-native English Speech Corpus}},
  year      = 2018,
  booktitle = {Proc. Interspeech}
}

@inproceedings{kominek2004cmu,
  title     = {The CMU Arctic speech databases},
  author    = {Kominek, John and Black, Alan W},
  booktitle = {Fifth ISCA workshop on speech synthesis},
  year      = {2004}
}

@inproceedings{he2024emilia,
  title        = {Emilia: An extensive, multilingual, and diverse speech dataset for large-scale speech generation},
  author       = {He, Haorui and Shang, Zengqiang and Wang, Chaoren and Li, Xuyuan and Gu, Yicheng and Hua, Hua and Liu, Liwei and Yang, Chen and Li, Jiaqi and Shi, Peiyang and others},
  booktitle    = {2024 IEEE Spoken Language Technology Workshop (SLT)},
  pages        = {885--890},
  year         = {2024},
  organization = {IEEE}
}

@inproceedings{lee2022direct,
  title     = {Direct Speech-to-Speech Translation With Discrete Units},
  author    = {Lee, Ann  and
               Chen, Peng-Jen  and
               Wang, Changhan  and
               Gu, Jiatao  and
               Popuri, Sravya  and
               Ma, Xutai  and
               Polyak, Adam  and
               Adi, Yossi  and
               He, Qing  and
               Tang, Yun  and
               Pino, Juan  and
               Hsu, Wei-Ning},
  booktitle = {Proc. ACL},
  year      = {2022}
}

@article{felps2009foreign,
  title   = {Foreign accent conversion in computer assisted pronunciation training},
  author  = {Felps, Daniel and Bortfeld, Heather and Gutierrez-Osuna, Ricardo},
  journal = {Speech communication},
  volume  = {51},
  number  = {10},
  pages   = {920--932},
  year    = {2009}
}

@inproceedings{turk2002subband,
  title     = {Subband based voice conversion.},
  author    = {T{\"u}rk, Oytun and Arslan, Levent M},
  booktitle = {Proc. Interspeech},
  pages     = {289--292},
  year      = {2002}
}

@inproceedings{sun2016personalized,
  title     = {Personalized, Cross-Lingual TTS Using Phonetic Posteriorgrams.},
  author    = {Sun, Lifa and Wang, Hao and Kang, Shiyin and Li, Kun and Meng, Helen M},
  booktitle = {Proc. Interspeech},
  pages     = {322--326},
  year      = {2016}
}

@inproceedings{zhao2018icassp,
  title     = {Accent conversion using phonetic posteriorgrams},
  author    = {Guanlong, Zhao and Sinem, Sonsaat and John, Levis and Evgeny, Chukharev-Hudilainen and Ricardo, Gutierrez-Osuna},
  pages     = {5314--5318},
  year      = {2018},
  booktitle = {Proc. ICASSP}
}

@article{li2020improving,
  title   = {Improving accent conversion with reference encoder and end-to-end text-to-speech},
  author  = {Li, Wenjie and Tang, Benlai and Yin, Xiang and Zhao, Yushi and Li, Wei and Wang, Kang and Huang, Hao and Wang, Yuxuan and Ma, Zejun},
  journal = {arXiv preprint arXiv:2005.09271},
  year    = {2020}
}

@inproceedings{quamer2022zero,
  title     = {Zero-Shot Foreign Accent Conversion without a Native Reference},
  author    = {Quamer, Waris and Das, Anurag and Levis, John and Chukharev-Hudilainen, Evgeny and Gutierrez-Osuna, Ricardo},
  booktitle = {Proc. Interspeech},
  pages     = {4920--4924},
  year      = {2022}
}

@inproceedings{liu2020end,
  title     = {End-to-end accent conversion without using native utterances},
  author    = {Liu, Songxiang and Wang, Disong and Cao, Yuewen and Sun, Lifa and Wu, Xixin and Kang, Shiyin and Wu, Zhiyong and Liu, Xunying and Su, Dan and Yu, Dong and others},
  booktitle = {Proc. ICASSP},
  pages     = {6289--6293},
  year      = {2020}
}

@inproceedings{zhao2019foreign,
  title     = {Foreign Accent Conversion by Synthesizing Speech from Phonetic Posteriorgrams.},
  author    = {Zhao, Guanlong and Ding, Shaojin and Gutierrez-Osuna, Ricardo},
  booktitle = {Proc. Interspeech},
  pages     = {2843--2847},
  year      = {2019}
}

@article{ding2022accentron,
  title   = {Accentron: Foreign accent conversion to arbitrary non-native speakers using zero-shot learning},
  author  = {Ding, Shaojin and Zhao, Guanlong and Gutierrez-Osuna, Ricardo},
  journal = {Computer Speech \& Language},
  volume  = {72},
  pages   = {101302},
  year    = {2022}
}

@article{zhao2021converting,
  title     = {Converting foreign accent speech without a reference},
  author    = {Zhao, Guanlong and Ding, Shaojin and Gutierrez-Osuna, Ricardo},
  journal   = {TASLP},
  volume    = {29},
  pages     = {2367--2381},
  year      = {2021},
  publisher = {IEEE}
}

@inproceedings{nguyen2022accent,
  title     = {Accent Conversion using Pre-trained Model and Synthesized Data from Voice Conversion.},
  author    = {Nguyen, Tuan-Nam and Pham, Ngoc-Quan and Waibel, Alexander},
  booktitle = {Proc. Interspeech},
  pages     = {2583--2587},
  year      = {2022}
}

@inproceedings{jin2023voice,
  title     = {Voice-preserving zero-shot multiple accent conversion},
  author    = {Jin, Mumin and Serai, Prashant and Wu, Jilong and Tjandra, Andros and Manohar, Vimal and He, Qing},
  booktitle = {Proc. ICASSP},
  year      = {2023}
}

@article{zhou2023tts,
  title     = {Tts-guided training for accent conversion without parallel data},
  author    = {Zhou, Yi and Wu, Zhizheng and Zhang, Mingyang and Tian, Xiaohai and Li, Haizhou},
  journal   = {Signal Processing Letters},
  volume    = {30},
  pages     = {533--537},
  year      = {2023},
  publisher = {IEEE}
}

@inproceedings{chen2024transfer,
  title     = {Transfer the linguistic representations from TTS to accent conversion with non-parallel data},
  author    = {Chen, Xi and Pei, Jiakun and Xue, Liumeng and Zhang, Mingyang},
  booktitle = {Proc. ICASSP},
  year      = {2024}
}

@inproceedings{bai2024diffusion,
  title     = {Diffusion-Based Method with TTS Guidance for Foreign Accent Conversion},
  author    = {Bai, Qibing and Wang, Shuai and Liu, Zhijun and Zhang, Mingyang and Rao, Wei and Wang, Yannan and Li, Haizhou},
  booktitle = {Proc. ISCSLP},
  pages     = {284--288},
  year      = {2024}
}

@inproceedings{bai2025accent,
  title     = {Accent Normalization Using Self-Supervised Discrete Tokens with Non-Parallel Data},
  author    = {Qibing Bai and Sho Inoue and Shuai Wang and Zhongjie Jiang and Yannan Wang and Haizhou Li},
  year      = {2025},
  booktitle = {Interspeech 2025},
  pages     = {1618--1622}
}

@inproceedings{nguyen24syndata4genai,
  title     = {Accent conversion using discrete units with parallel data synthesized from controllable accented TTS},
  author    = {Tuan-Nam Nguyen and Quan Pham and Alexander Waibel},
  year      = {2024},
  booktitle = {Synthetic Data’s Transformative Role in Foundational Speech Models},
  pages     = {51--55}
}

@inproceedings{nguyen2024improving,
  title     = {Improving Pronunciation and Accent Conversion through Knowledge Distillation And Synthetic Ground-Truth from Native TTS},
  author    = {Nguyen, Tuan Nam and Akti, Seymanur and Pham, Ngoc Quan and Waibel, Alexander},
  booktitle = {ICASSP},
  year      = {2025}
}

@inproceedings{zhang2025vevo,
  title     = {Vevo: Controllable Zero-Shot Voice Imitation with Self-Supervised Disentanglement},
  author    = {Xueyao Zhang and Xiaohui Zhang and Kainan Peng and Zhenyu Tang and Vimal Manohar and Yingru Liu and Jeff Hwang and Dangna Li and Yuhao Wang and Julian Chan and Yuan Huang and Zhizheng Wu and Mingbo Ma},
  booktitle = {ICLR},
  year      = {2025}
}

@article{bai2026cosyaccent,
  title   = {CosyAccent: Duration-Controllable Accent Normalization Using Source-Synthesis Training Data},
  author  = {Bai, Qibing and Shi, Shuhao and Wang, Shuai and Ju, Yukai and Wang, Yannan and Li, Haizhou},
  journal = {Proc. ICASSP 2026},
  year    = {2026}
}

@article{halychanskyi2025fac,
  title   = {FAC-FACodec: Controllable Zero-Shot Foreign Accent Conversion with Factorized Speech Codec},
  author  = {Halychanskyi, Yurii and Churchwell, Cameron and Wen, Yutong and Kindratenko, Volodymyr},
  journal = {Proc. ICASSP 2026},
  year    = {2026}
}

@inproceedings{jia2024convert,
  title     = {Convert and Speak: Zero-shot Accent Conversion with Minimum Supervision},
  author    = {Zhijun Jia and Huaying Xue and Xiulian Peng and Yan Lu},
  booktitle = {Multimedia},
  year      = {2024}
}

@article{du2024cosyvoice,
  title   = {Cosyvoice 2: Scalable streaming speech synthesis with large language models},
  author  = {Du, Zhihao and Wang, Yuxuan and Chen, Qian and Shi, Xian and Lv, Xiang and Zhao, Tianyu and Gao, Zhifu and Yang, Yexin and Gao, Changfeng and Wang, Hui and others},
  journal = {arXiv preprint arXiv:2412.10117},
  year    = {2024}
}

@article{liu2024controllable,
  author  = {Liu, Rui and Sisman, Berrak and Gao, Guanglai and Li, Haizhou},
  title   = {Controllable Accented Text-to-Speech Synthesis With Fine and Coarse-Grained Intensity Rendering},
  year    = {2024},
  volume  = {32},
  issn    = {2329-9290},
  doi     = {10.1109/TASLP.2024.3378110},
  journal = {IEEE/ACM Transactions on Audio, Speech, and Language Processing},
  month   = apr,
  pages   = {2188–2201}
}

@inproceedings{lipman2023flow,
  title     = {Flow Matching for Generative Modeling},
  author    = {Yaron Lipman and Ricky T. Q. Chen and Heli Ben-Hamu and Maximilian Nickel and Matthew Le},
  booktitle = {ICLR},
  year      = {2023}
}

@inproceedings{peebles2023scalable,
  title     = {Scalable diffusion models with transformers},
  author    = {Peebles, William and Xie, Saining},
  booktitle = {Proc. ICCV},
  pages     = {4195--4205},
  year      = {2023}
}

@inproceedings{ho2021classifierfree,
  title     = {Classifier-Free Diffusion Guidance},
  author    = {Jonathan Ho and Tim Salimans},
  booktitle = {NeurIPS 2021 Workshop on Deep Generative Models and Downstream Applications},
  year      = {2021}
}

@inproceedings{xinyuan2025scalable,
  title     = {Scalable Controllable Accented TTS},
  author    = {Xinyuan, Henry Li and Cai, Zexin and Garg, Ashi and Duh, Kevin and Garc{\'\i}a-Perera, Leibny Paola and Khudanpur, Sanjeev and Andrews, Nicholas and Wiesner, Matthew},
  booktitle = {Proc. ASRU 2025},
  year      = {2025}
}

@inproceedings{wang2023nonparallel,
  title     = {Non-parallel Accent Transfer based on Fine-grained Controllable Accent Modelling},
  author    = {Wang, Linqin  and
               Yu, Zhengtao  and
               Yang, Yuanzhang  and
               Gao, Shengxiang  and
               Mao, Cunli  and
               Huang, Yuxin},
  booktitle = {EMNLP 2023},
  year      = {2023},
  publisher = {Association for Computational Linguistics},
  pages     = {9288--9298}
}

@inproceedings{van2017vq,
  author    = {van den Oord, Aaron and Vinyals, Oriol and Kavukcuoglu, Koray},
  title     = {Neural discrete representation learning},
  year      = {2017},
  booktitle = {Proceedings of the 31st International Conference on Neural Information Processing Systems},
  pages     = {6309–6318},
  numpages  = {10}
}

@article{su2024roformer,
  title   = {Roformer: Enhanced transformer with rotary position embedding},
  author  = {Su, Jianlin and Ahmed, Murtadha and Lu, Yu and Pan, Shengfeng and Bo, Wen and Liu, Yunfeng},
  journal = {Neurocomputing},
  volume  = {568},
  pages   = {127063},
  year    = {2024}
}

@inproceedings{mcauliffe2017montreal,
  title     = {Montreal forced aligner: Trainable text-speech alignment using kaldi.},
  author    = {McAuliffe, Michael and Socolof, Michaela and Mihuc, Sarah and Wagner, Michael and Sonderegger, Morgan},
  booktitle = {Proc. Interspeech},
  pages     = {498--502},
  year      = {2017}
}

@article{ravillion2020comparison,
  title  = {A comparison of Best-Worst Scaling and Rating Scale for timbre characterisation},
  author = {Ravillion, Aliette Marie Veronique},
  year   = {2020}
}

@inproceedings{churchwell2024high,
  title     = {High-fidelity neural phonetic posteriorgrams},
  author    = {Churchwell, Cameron and Morrison, Max and Pardo, Bryan},
  booktitle = {ICASSP 2024 Workshop on Explainable Machine Learning for Speech and Audio},
  year      = {2024}
}

\end{document}